\documentclass[reprint,bibnotes,aps,prl,floatfix,superscriptaddress,longbibliography]{revtex4-1}
\usepackage{graphicx}
\usepackage{amsmath}
\usepackage{amsxtra}
\usepackage{amssymb}
\usepackage{dcolumn}
\usepackage{float}
\usepackage{bm}
\usepackage{verbatim}
\usepackage[breaklinks=true,colorlinks,citecolor=blue,linkcolor=blue,urlcolor=blue]{hyperref}

\def\bc{\begin{equation}}
\def\ec{\end{equation}}
\def\bea{\begin{eqnarray}}
\def\eea{\end{eqnarray}}
\DeclareMathAlphabet\mathbfcal{OMS}{cmsy}{b}{n}

\begin{document}

\title{A novel undamped gapless plasmon mode in tilted type-II Dirac semimetal}
\author{Krishanu Sadhukhan}
\affiliation{Department of Physics, Indian Institute of Technology , Kanpur 208016, India}
\author{Antonio Politano}
\affiliation{Dipartimento di Scienze Fisiche e Chimiche (DSFC), Universit\`a dell'Aquila, Via Vetoio 10, I-67100 L'Aquila, Italy}
\author{Amit Agarwal}
\email{amitag@iitk.ac.in}
\affiliation{Department of Physics, Indian Institute of Technology , Kanpur 208016, India}

\date{\today}

\begin{abstract}
We predict the existence of a novel long-lived gapless plasmon mode in a type-II Dirac semimetal (DSM). 
This gapless mode arises from the out-of-phase oscillations of the density fluctuations in the electron and the hole pockets of a type-II DSM. 
It originates beyond a critical wave-vector along the direction of the tilt axis, owing to the momentum separation of the electron and hole pockets. A similar out-of-phase plasmon mode arises in other multi-component charged fluids as well, but generally it is Landau damped and lies within the particle-hole continuum. In the case of a type-II DSM, 
the open Fermi surface prohibits low-energy finite momentum single-particle excitations, creating a `gap' in the particle-hole continuum. The gapless plasmon mode lies within this particle-hole continuum gap and, thus, it is protected from Landau damping.  

\end{abstract}

\maketitle
\begin{figure}
\includegraphics[width=.99\linewidth]{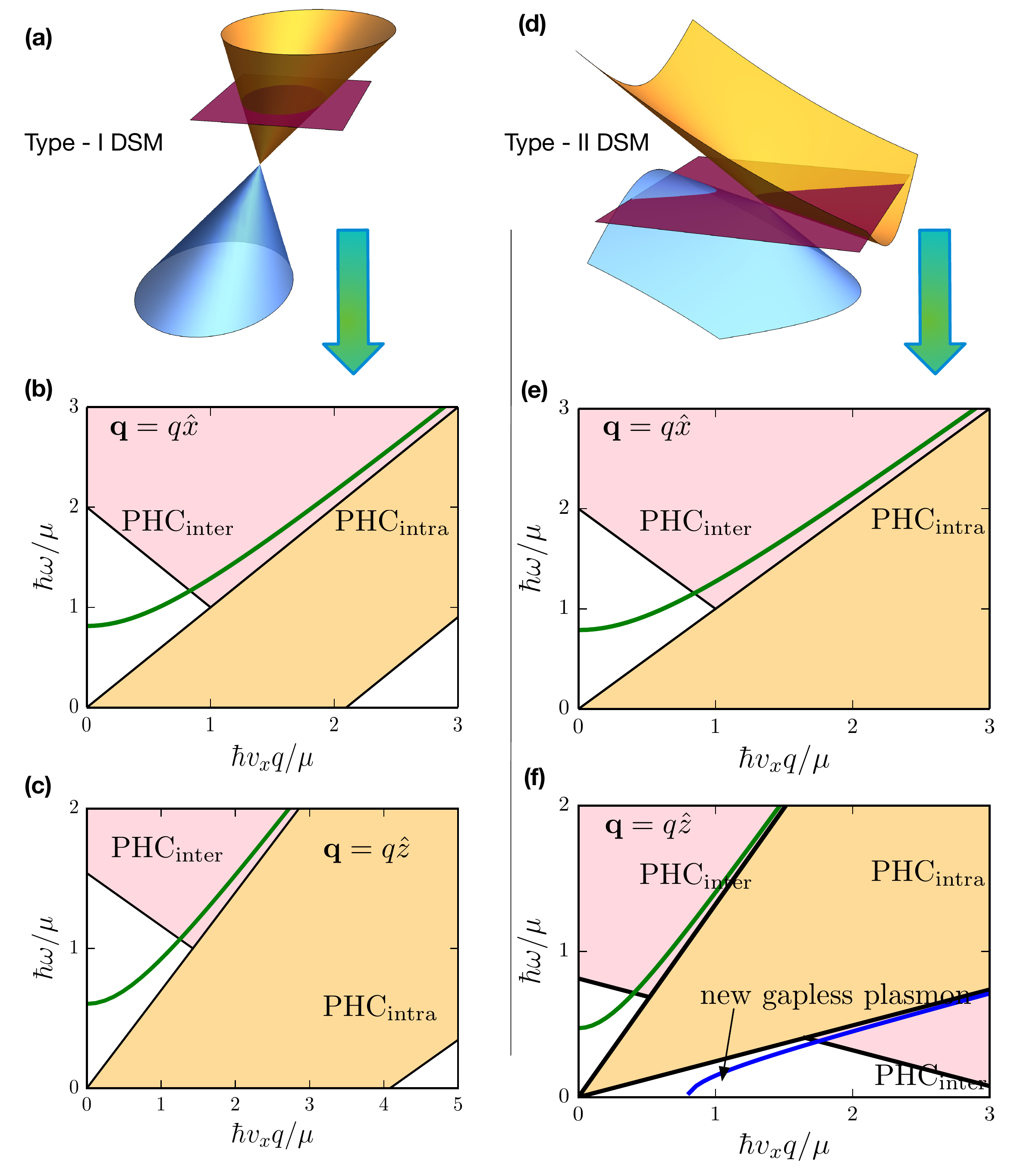}
\caption{Schematic of the  band structure ($\varepsilon_{\bf k}$) of DSM node for a (a) type-I and (d) type-II DSM. 
The closed elliptic Fermi surface with an electron  pocket in a type-I DSM, and the open hyperbolic Fermi surface with an electron and a hole pocket in type-II DSM is evident. 
Panels (b) and (c) depict the PHC (both inter- and intra-band) along with the gapped plasmon mode for a type-I DSM for $q$ along the $\hat x$ and $\hat z$ direction, respectively. Panels (e) and (f) show the PHC (both inter- and intra-band) and the plasmon modes of type-II DSM along the $\hat x$ and $\hat z$ direction, respectively. The particle-hole gap and the novel gapless plasmon mode (blue line) are also shown. Here, we have chosen $v_x=v_y=v_F=0.65\times10^6$m/s, $v_z=0.35\times10^6$m/s and $v_t=0.6(1.46)v_z$ for a type-I (type-II) DSM. 
\label{fig1}} 
\end{figure}

The topological semimetal state in crystalline solids allows for the existence of relativistic quasiparticles, which have no analogue in the standard model \cite{Bansil2016,Ashvin2018}. 
Protected by crystalline symmetries, the anisotropic tilted type-I and type-II Dirac (DSM) and Weyl semimetal (WSM) phases \cite{Burkov2018}, which violate the Lorentz invariance, are  examples of this class of materials\cite{Yang2014,Soluyanov2015,PhysRevLett.119.026404}. For a tilted DSM, the electronic dispersion is a sum of `potential' and `kinetic' 
terms, $E_{\bf k}=U_{\bf k}\pm T_{\bf k}$, where both the terms vanish at the fourfold degenerate  (including spin) Dirac point \cite{Yan2017,PtTe2PRL}. The first term is odd along a specific direction of $\bf k$ (the `tilt' direction), while the second term has the usual form of an anisotropic massless Dirac cone. The DSM phase is type-II if the Dirac cone is tilted over or $U_{\bf k}>T_{\bf k}$ along the tilt direction, else it is classified as type-I DSM including the case of isotropic DSM without tilt. 
Experimental realizations of a type-I DSM include Na$_3$Bi\cite{Wang2012,Liu2014} and Cd$_3$As$_2$\cite{Wang2013,Liu2014_2} among others. Type-II DSM phase has been identified in PtTe$_2$\cite{Yan2017,Nmat_TMD,PtTe2PRL}, PtSe$_2$\cite{PhysRevB.96.125102}, PdTe$_2$ \cite{PhysRevLett.119.016401,Nmat_TMD,PhysRevB.96.041201} among others. 
The Fermi surface in a type-I DSM is an ellipsoid enclosing a single type of carrier pocket (either electron or hole), with a vanishing density of states (DOS) at the Dirac point. Contrarily, the Fermi surface of a type-II DSM  is a hyperboloid with {\it open} electron and hole pockets along the tilt axis, with a finite DOS at the Dirac point, as shown in Fig.~\ref{fig1}(a)-(d). The presence of both types of carriers at the Fermi energy in a type-II DSM leads to several interesting magneto-transport and optical properties \cite{2018arXiv180803646D,PhysRevLett.119.176804}. Here, we explore collective density excitations in a type-II DSM \cite{PtTe2PRL,PhysRevB.99.045414} and predict the existence of a novel undamped gapless plasmon mode. 

The presence of both electron and hole pockets at the Fermi energy in a type-II DSM suggests the possibility of two plasmon modes, related to the `in-phase' and the 
`out-of-phase' oscillation of the density deviations in the two electron fluids. The in-phase oscillation leads to the normal gapped plasmon mode in three-dimensional systems \cite{PhysRevB.91.205426}, 
while the out-of-phase oscillations generally lead to gapless plasmon mode. Similar gapless plasmon mode has been reported in several two-component systems, including spatially separated 2D electron liquids \cite{PhysRevB.23.805,PhysRevB.82.195428}, bilayer graphene \cite{PhysRevB.82.195428}, and spin-polarized systems \cite{PhysRevB.90.155409} among others. However, the out-of-phase gapless mode is generally damped, as it lies within the particle-hole continuum (PHC) \cite{PhysRevB.23.805,PhysRevB.82.195428,PhysRevB.82.195428,PhysRevB.90.155409}. In contrast to this, we show that in type-II DSM the out-of-phase plasmon mode is undamped. 

Our demonstration of this novel gapless plasmon mode in type-II DSM is based on hydrodynamic theory along with exact analytical calculation of the density response function. 
The gapless plasmon mode appears beyond a critical wave-vector on account of the momentum separation of the electron and the hole pockets along the tilt axis.  
Interestingly, the hyperboloidal open Fermi surface in a type-II DSM prohibits particle-hole excitations for vanishing energies and finite wave-vectors along the tilt axis, creating a `PHC gap' in the single-particle excitation spectrum. The predicted gapless plasmon mode lies within this PHC gap, protected from Landau damping, and is long-lived for small energies. Additionally, we also explore the impact of tilt on the three-dimensional gapped `in-phase' plasmon mode, which has an anisotropic plasmon gap depending on the direction of approach for the long-wavelength limit.We also elucidate the experimental requirements to detect this novel plasmon.

In tilted DSM, the oppositely tilted Dirac nodes generally appear in pairs on different $k-$points \cite{Yang2014} located on the $C_3$ rotation axis (usually the $\Gamma-A$ direction, chosen to be the $\hat z$-axis). 
For simplicity, we consider tilted DSM hosting only one pair of Dirac nodes tilted along the $\hat z$-axis, as in PtSe$_2$ \cite{PhysRevB.96.125102}, PdTe$_2$ \cite{PhysRevLett.119.016401,PhysRevB.96.041201}, and PtTe$_2$\cite{Yan2017,PtTe2PRL} and other materials \cite{Wang2012,Liu2014,Wang2013,Liu2014_2,Singh2018}. The general low-energy Hamiltonian for each nodes (excluding the spin degeneracy) is,
\bc\label{hamiltonian}
\mathcal{H_\chi}=\hbar[\chi v_t k_z\sigma_0 + v_x k_x \sigma_x +v_y k_y \sigma_y +v_z k_z \sigma_z]~.
\ec
Here, $v_i$ ($i = x,y,z$) denotes the Fermi velocity along $i$ direction,  $v_t$ is the tilt velocity, $\chi=\pm1$ is the node index, $\sigma_i$  are the Pauli spin matrices and $\sigma_0$ is the ($2  \times 2$) unit matrix. 
Depending on the value of the tilt parameter $\beta = |v_t|/|v_z|$, the boundaries of Dirac cone along the $z$-axis have either opposite or same sign of their slopes resulting in a type-I ($\beta <1$) or a type-II DSM ($\beta >1$). 
The energy dispersion, $\varepsilon^\chi_{\lambda{\bf k}}=\chi\hbar v_t k_z+\lambda\hbar\sqrt{v_x^2k_x^2+v_y^2k_y^2+v_z^2k_z^2}$, for a single node of both type-I and type-II DSM is shown in Fig.~\ref{fig1}(a) and (d), respectively. Here, $\lambda \pm 1$ is the band index denoting the conduction/valance band. 
The differing topology of 
the Fermi surfaces for a type-I and a type-II DSM along the tilt axis is evident. 
While the Fermi surface for each node of a type-I DSM is a closed ellipsoid, type-II DSM has a pair of open hyperboloid Fermi surfaces along the tilt axis, enclosing an electron and a hole pockets.

For a doped type-I DSM with a closed electron pocket,  the structure of the PHC is very similar to that in an isotropic DSM \cite{thakur2017,Thakur2018}. The intra-band 
single-particle transitions occur only for $q_i$ lying between between the $q_i=0$ and $q_i < \mu/(\hbar v_i)$  (for $\mu > 0$) lines [see Fig.~\ref{fig1}(b)-(c)]. 
Qualitatively, this also occurs in a type-II DSM, for ${\bf q}$ perpendicular to the tilt axis (in the ${\hat x}-{\hat y}$ plane) - see Fig.~\ref{fig1}(e). 

Along the tilt axis (${\bf q} = q {\hat z}$) in a type-II DSM, the open (hyperboloid) Fermi surface restricts all low-energy finite-$q$ intra-band transitions, causing the PHC to lie between the lines $\hbar \omega = (v_t \pm v_z) q$. 
More interestingly, the coexistence of an electron and a hole pocket in a type-II DSM results in low-energy inter-band transition for $q \ge q_{eh}$. 
Here, $q_{eh}=2\mu/[\hbar v_z (\beta^2-1)]$ quantifies the momentum separation between the electron and hole pockets for a fixed chemical potential ($\mu$). 
Both these combine to produce a PHC spectrum, which has a `PHC gap' in the low-energy finite-$q$ regime for $q < q_{eh}$ [see Fig.~\ref{fig1}(f)]. We refer the reader to Sec.~S2 of the SM\cite{Note1} for a thorough discussion on the PHC spectrum in tilted DSM. 
\footnote{Supplementary material includes 1) details of the Fermi surface and calculation and PHC, 2) analytical results for the non-interacting density-density response function, 3) the gapped plasmon mode in type-I and type-II DSM and 4) the dispersion of the gapless plasmon mode in type-II DSM.}.
Below, we show that this `PHC gap' hosts a novel gapless Dirac plasmon mode in type-II DSM. 

In order to understand the nature of the collective plasmon modes in tilted DSM, we start with the classical hydrodynamics approach \cite{GV}. First we generalize the 
classical hydrodynamics approach to  include 1) the anisotropic effective mass in DSM and 2) the presence of multiple charged fluids interacting with each other via the Coulomb interaction. 
In the continuum limit, the electronic density fluctuation for the $i$th electron liquid can be expressed as $\rho_i({\bf r},t)=\rho_{0i}+\rho_{1i}({\bf r},t)$, where $\rho_{0i} = e n_{0i}$ is the constant background density 
and $\rho_{1i}({\bf r},t)\ll \rho_{0i}$ is the small density deviation.  The corresponding electronic current density ${\bf
j}_i({\bf r},t)$ satisfies the local continuity equation: $\frac{\partial \rho_i}{\partial t}+\nabla \cdot{\bf j}_i=0$. For simplicity, we neglect damping, and assume the electron fluids to be incompressible. Then, the Euler-Lagrange equation of motion for the $i$th electron fluid is given by vector equation: ${\cal{M}}_i \partial_t {\bf j}_i = e^2 n_{0i} {\bf E}_i({\bf r}, t)$, ~where ${\cal M}_i$ is the effective mass matrix. Here, ${\bf E}_i({\bf r},t) = - \nabla_{\bf r} \phi^{\rm int} ({\bf r},t) $ is the effective electric field and  $\phi^{\rm int} = \sum_j \int d{\bf r}' e^2 \rho_{1j}/|{\bf r} - {\bf r}'|$. Using the continuity equation and doing simple matrix manipulations, we obtain the following equation for density fluctuations: 
$\partial_t^2 \rho_{1i} = e^2 n_{0i} \nabla_{\bf r} \cdot \left[{\cal M}_i^{-1}~\nabla_{\bf r} \phi^{\rm int} ({\bf r},t) \right]$.
For a system with $N$ electron fluids, these are $N$ coupled equations. Further simplification is made by assuming the effective mass tensor to be diagonal, and neglecting the spatial variation of ${\cal M}_i$ to obtain the long-wavelength results. Taking the Fourier transform, 
we obtain the following equation for the long-wavelength collective modes:  ${\rm det}|\omega^2 \openone_N - {\cal A}({\bf q})| = 0$, which, in principle, supports $N$ collective modes. 
Here, we have defined ${\cal A}_{lm} = n_{0l} V_{lm}({\bf q})~({\bf q} {\cal M}_{l} {\bf q})$, where $V_{lm}$ is the Fourier transform of the Coulomb interaction between the $l$ and $m$ electron liquids, and  ${\bf q} {\cal M}^{l} {\bf q} = 
\sum_{k=x,y,z} q_k^2/({\cal M}_{l})_{kk}$. For unscreened Coulomb interaction, $V({\bf q}) \equiv V_q = e^2/(\epsilon_0\epsilon_rq^2)$, with $\epsilon_r$ being the relative dielectric constant of the material.

In order to incorporate the anisotropic effective mass in DSM, we use the following definition of the inertial effective mass: $ m_i^\lambda = \hbar^2 {k_i}/(\partial_{k_i} E_{\bf k}^\lambda)$, with $E_{\bf k}^\lambda$ denoting the electronic dispersion. 
This reproduces the conventional effective mass for parabolic band systems, as well as the inertial (and cyclotron) mass in graphene: $m = \mu/v_F^2$.  
For the tilted DSM dispersion of a given node with $\mu >0$, we obtain
\bc\label{eff mass3}
m_{\{x,y,z\}}^{\lambda\chi}=\frac{\lambda}{|\lambda+\chi\beta\cos\theta_q|}\left\{\frac{\mu}{v_x^2},\frac{\mu}{v_y^2},\frac{\mu \cos \theta_{q}}{v_z^2(\chi\beta+\lambda\cos\theta_q)}\right\}.
\ec
Here, $\beta$ quantifies the angle of the tilt and we have defined $\cos \theta_q = 
v_z q_z/\sqrt{v_x^2 q_x^2 + v_y^2 q_y^2 + v_z^2 q_z^2}$. As $\beta \to 0$, Eq.~\eqref{eff mass3} yields $
m_i = \mu/v_i^2$. 
In a type-I DSM, the conduction band electron effective mass is always positive. However, 
in a type-II DSM, even for a given node with $\mu >0$ we have $\lambda = \pm 1$, resulting in different effective masses for the electron and the hole pockets.

A type-I DSM node hosts a single charged fluid for any $\mu$ (electron liquid for $\mu > 0$), with an anisotropic mass. Thus, we find the conventional 3D gapped Dirac plasmons 
with anisotropic dispersion and different plasmon gap depending on the direction of approach to the $q \to 0$ limit: 
\bc\label{plDSM1a}
\omega_{\rm pl}^2 \approx  \frac{2n_1e^2}{\epsilon_0\epsilon_r\mu} \times
\begin{cases}
{v_z^2(1+\beta^2)}~ +  {\cal O}(q^2)~  & \text{for}~{\bf q} = q {\hat z},\\
{v_x^2} f^2(\phi_q)~ + {\cal O}(q^2) & \text{for}~{\bf q} = q {\hat n}_{x-y}~\\
\end{cases}.
\ec
Here, $n_1=2\mu^3/[3\pi^2\hbar^3v_xv_yv_z(1-\beta^2)^{2}]$ is the total electron density (per node for $\mu >0$) in a type-I DSM including spin. 
Additionally, $\phi_q = \tan^{-1}(q_y/q_x)$ is the azimuthal angle of ${\bf q}$ and $f^2(\phi_q)=\cos^2\phi_q+(v_y^2/v_x^2)\sin^2\phi_q$. 
Equation~\eqref{plDSM1a} reproduces the known result for the isotropic case without any tilt. 
The plasmon dispersion for a type-I DSM, along with the corresponding PHC, is shown in Fig.~\ref{fig1}(b) and (c). 

The case of a type-II DSM hosting an electron and a hole pocket, separated in the momentum space along the $k_z$ axis, is more interesting. Using the hydrodynamic theory for multi-component fluids developed above, the plasmon dispersion along ${\bf q} = q {\hat z}$ is given by the roots of, 
\bc\label{plDSM2aa}
\omega^2=\frac{n_{2}v_z^2q^2}{\mu}\Big(v_q(\beta+1)^2-v_{|q+q_{eh}|}(\beta-1)^2\Big)~.
\ec 
Here, $n_2 \approx {\mu\mathcal{E}_{\rm max}^2}/{(12\pi^2\beta\hbar^3v_xv_yv_z)}$ denotes the cutoff ($\mathcal{E}_{\rm max}$) dependent electron density for each node (with spin) of a type-II DSM. Equation~\eqref{plDSM2aa} permits two solutions. One of these is the conventional gapped plasmon mode in the limit $q_z \to 0$, whose dispersion is given by, 
\bc \label{plDSM2bb}
\omega_{\rm pl}^2\approx\frac{n_2e^2}{\epsilon_0\epsilon_r\mu} v_z^2(1+\beta)^2 + {\cal O}(q^2),~\hspace{.4cm}\text{for}~{\bf q} = q {\hat z}. 
\ec
In the $x-y$ plane, the plasmon gap for a type-II DSM is identical to that of type-I DSM in Eq.~\eqref{plDSM1a}, with the replacement $n_1 \to n_2$. 

Interestingly, Eq.~\eqref{plDSM2aa} permits another gapless solution ($\omega \to 0$) beyond a critical wave-vector, $q >q_c \equiv {\mu\beta}/{2\hbar v_z(\beta-1)}$. 
Expanding the right hand side of Eq.~\eqref{plDSM2aa} around $q_c$, we find the dispersion of the novel gapless plasmon mode ($\omega_{\rm npl}$) to be 
\bc
\omega_{\rm npl}^2 \approx \frac{8 n_2 e^2}{\epsilon_0\epsilon_r} \frac{\hbar v_z^3 (1+\beta)^2}{\mu^2 \beta} (q-q_c),~~\hspace{.3cm}\text{for}~{\bf q} = q {\hat z}.
\ec
More remarkably, this low-energy finite $q$ plasmon mode lies in the `PHC gap' arising from the open nature of the Fermi-surface along the $q_z$ direction.
Consequently, this novel mode remains undamped with a large quality factor till it enters the PHC [see Fig.~\ref{fig1}(d)].  
Physically, this novel mode arises from the out-of-phase {\it intra-node} density oscillations of the electron and hole fluids in a type-II DSM   
\footnote{This can be shown explicitly in the $q\to 0$ limit, using the matrix form of the coupled hydrodynamic equations for the electron and hole density deviations}.

Going beyond the hydrodynamic theory, we now demonstrate the existence of this novel undamped plasmon mode 
by calculating the density response function explicitly. The electron-electron interaction will be treated within the random phase approximation (RPA), which is well known to describe plasmons in several systems. 
The non-interacting density response (or Lindhard) function of a single Dirac node is given by \cite{GV,PhysRevB.83.115135,PhysRevB.91.205426,Ghosh2017,Thakur2018,PhysRevB.96.035410},
\bea \label{response_func}
\Pi^{\text{NI}}_\chi ({\bf q},\omega)&=&\frac{g_s}{V} \sum_{{\bf k}, \lambda, \lambda'}\frac{f({\varepsilon_{\lambda{\bf k}}^\chi})-f({\varepsilon_{\lambda'{\bf k+q}}^\chi})}{\hbar\omega^+ + \varepsilon_{\lambda{\bf k}}^\chi- \varepsilon_{\lambda'{\bf k+q}}^\chi}
F_{\lambda \lambda'}({\bf k},{\bf q}).
\eea
Here, $V$ is the volume, $\omega^+ = \omega + i \eta$ with $\eta \to 0$, and  $F_{\lambda \lambda'}({\bf k},{\bf q})$ is the orbital overlap function. The Fermi function $f(\varepsilon)=[1+\exp(\frac{\varepsilon-\mu}{k_bT})]^{-1}$ acts as a step function at $T=0$. The total density response function includes both nodes: $\Pi^{\text{NI}}=\Pi^{\text{NI}}_++\Pi^{\text{NI}}_-$. The analytical calculation of $\Pi^{\text{NI}}$ for both type-I and type-II DSM is detailed in Sec.~S3-S4 of the SM \cite{Note1}. 
The density response function for an interacting electron fluid, within RPA, is given by %
\bc \label{Piint}
\Pi^{\rm RPA}({\bf q},\omega) = \Pi^{\text{NI}}({\bf q},\omega)/\epsilon({\bf q},\omega)~. 
\ec %
Here, $\epsilon({\bf q},\omega)\equiv 1-V_q \Pi^{\text{NI}}({\bf q},\omega)$ is the dynamical dielectric function. 
The plasmon dispersion and damping constant, $\omega_{\rm pl}({\bf q}) - i \gamma_{\rm pl}({\bf q})$, can now be obtained from the poles of $\Pi^{\rm RPA}({\bf q},\omega)$, or alternately from the complex roots of $\epsilon({\bf q},\omega)=0$.
For small damping rate, an expansion of $\epsilon({\bf q},\omega)$ around $\omega_{\rm pl}$, yields $\gamma_{\rm pl} = 
\frac{{\rm Im}[\epsilon]}{\partial_\omega {\rm Re}[\epsilon]} \big|_{\omega_{\rm pl}} > 0$. For a stable plasmon mode $\gamma_{\rm pl} > 0$. 
\begin{figure}
\includegraphics[width=1.0\linewidth]{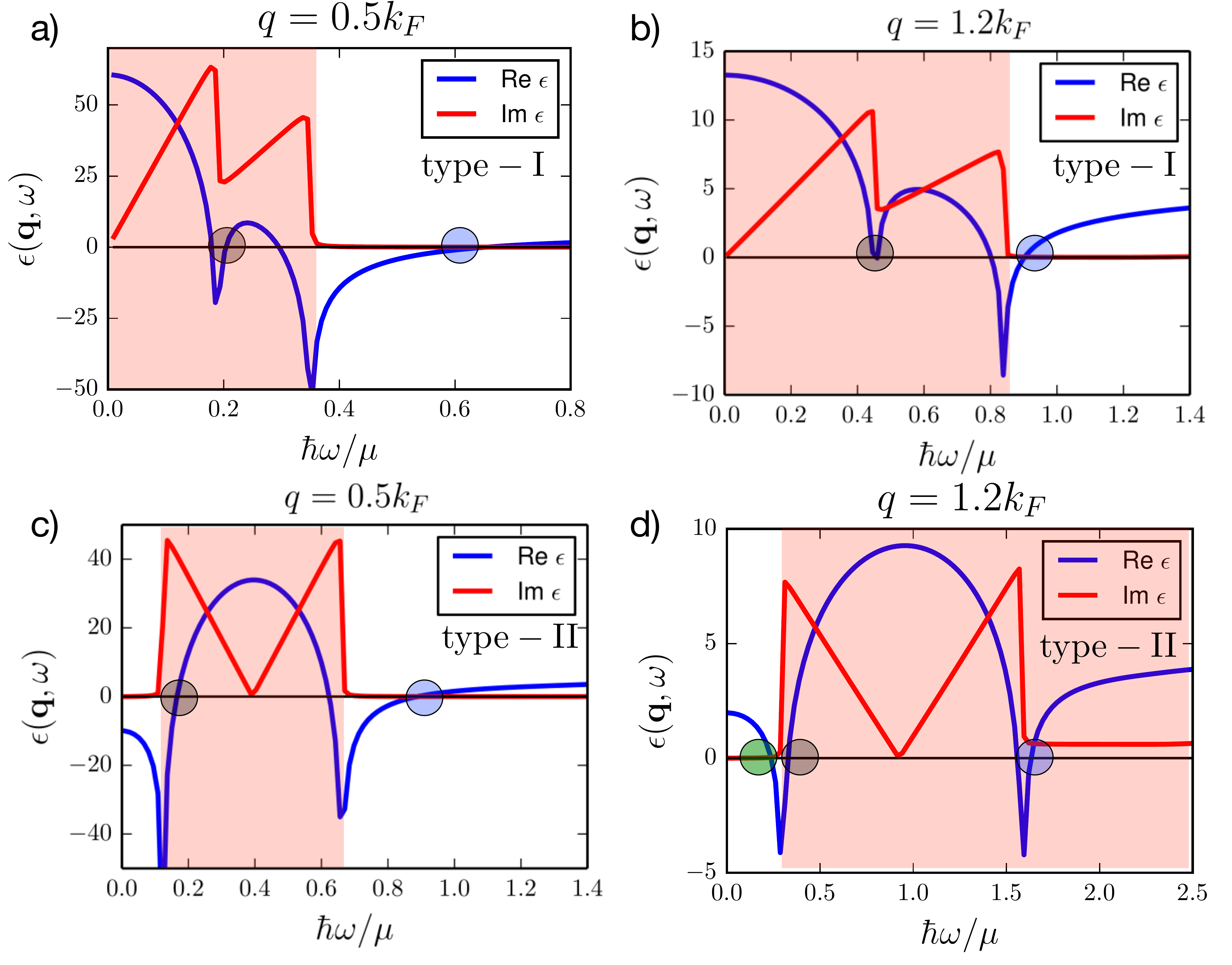}
\caption{The real and imaginary parts of the dielectric function, $\epsilon({\bf q},\omega)$ for two different values of $q$ along the tilt axis, for type-I DSM in (a) and (b), and for type-II DSM in (c) and (d). The shaded region depicts the PHC, and the circles denote the zeros of $\epsilon({\bf q},\omega)$ corresponding to collective excitations. The blue circle is the regular gapped plasmon, while the grey circle is the highly damped mode arising from the out-of-phase oscillation of the electron fluids in different nodes. The green circle in (d) is the novel gapless plasmon mode arising from the out-of-phase oscillations of the intra-node electron/hole pockets in a type-II DSM. Other parameters are dientical to that of Fig.~\ref{fig1}.
\label{fig2}} 
\end{figure}

The dielectric function for both type-I and type-II DSM is shown in Fig.~\ref{fig2} for ${\bf q} (= q{\hat z})$ along the tilt axis. Panels (a) and (b) for a type-I DSM show the existence of two stable plasmon modes (same sign of ${\rm Im}[\epsilon]$ and $\partial_\omega {\rm Re}[\epsilon]$, at the ${\rm Re}[\epsilon] = 0$ crossings). The rightmost root of ${\rm Re}[\epsilon] = 0$ with a vanishingly small ${\rm Im} [\epsilon]$ is the gapped 3D Dirac plasmon mode \cite{PtTe2PRL, PhysRevB.99.121401}.
The other root of ${\rm Re}[\epsilon] = 0$ is the damped plasmon mode. It lies in the PHC and corresponds to the out-of-phase oscillations of the electron fluids in different Dirac nodes (see Fig.~S2 in SM\cite{Note1}). 
The dielectric function for a type-II DSM is shown in panels (c) and (d) of Fig.~\ref{fig2} for small and large $q$ along the tilt axis, respectively. For small $q$ in panel (c), there are two stable collective modes: the 3D gapped Dirac plasmon, and the Landau damped mode originating from the out-of-phase {\it inter-node} density oscillations as discussed above. This changes drastically for large $q_z$ in panel (d), with the emergence of the novel undamped gapless plasmon mode for $q_z > q_c$ at low energies - qualitatively consistent with the 
predictions of the multi-fluid hydrodynamic theory. Physically, this mode corresponds to out-of-phase {\it intra-node} density oscillations in the electron and hole pockets in type-II DSM \cite{Note2}.  

\begin{figure}
\includegraphics[width=1.0\linewidth]{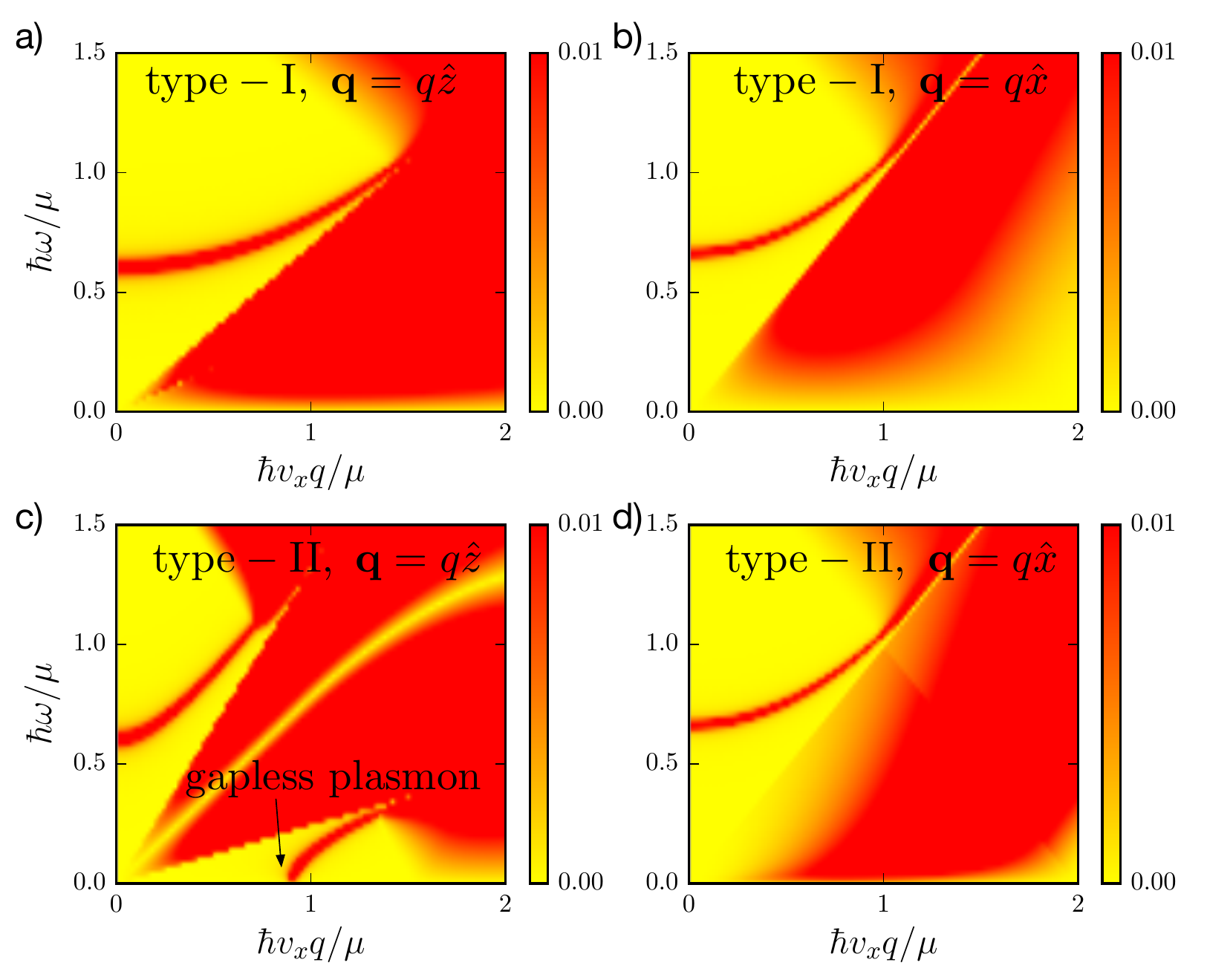}
\caption{The RPA loss function, ${\cal{E}}_{\rm loss}({\bf q},\omega)$, for a type-I [(a) and (b)] and type-II DSM [(c) and (d)] along different directions. 
The anisotropic band-structure results in an anisotropic plasmon gap. 
The `PHC gap' in the low-energy finite-$q$  loss function, and the existence of the novel undamped plasmon mode for ${\bf q}$ along the tilt axis is evident in (c). Other parameters are identical to that of Fig.~\ref{fig1}.
\label{fig3}} 
\end{figure}

Experimentally, plasmon resonances also appear as peaks in the momentum resolved electron energy loss spectrum (EELS), which probes the loss function: ${\cal{E}}_{\rm loss}({\bf q},\omega) =-{\rm Im}[1/\epsilon({\bf q},\omega)]$.
The loss function for the type-I and type-II DSMs, for ${\bf q}$ along different directions, is shown in Fig.~\ref{fig3}. For type-I DSM the gapped Dirac plasmon mode with an anisotropic energy gap is evident in Fig.~\ref{fig3} (a) and (b), for ${\bf q} = q {\hat z}$ and ${\bf q} = q {\hat x}$, respectively.  The Landau-damped, out-of-phase inter-node plasmon mode lies within the PHC spectrum in panel (a), and it is not clearly visible. 

The RPA loss function, ${\cal{E}}_{\rm loss}({\bf q},\omega)$, for a type-II DSM is shown in Fig.~\ref{fig3} (c) and (d), for ${\bf q} = q {\hat z}$ and ${\bf q} = q {\hat x}$, respectively. Panel (c) clearly highlights the 1) `PHC gap' in the low-energy but finite-$q$ loss spectrum for $q$ along the tilt axis, and 2) the 
existence of the novel undamped gapless plasmon mode (for $q>q_c$).
Our analytical calculations for the density response function coupled with RPA yield the critical wave-vector (for $\beta >1$) to be 
\bc
q_c \approx \frac{\mu}{\hbar v_z}\sqrt{\frac{2 g_s \alpha_{\rm fine}}{\pi}G(\beta)},~\hspace{.2cm}\text{where} \hspace{.2cm} G(\beta)=\beta\ln\frac{\beta+1}{\beta-1}-2, 
\ec
and $\alpha_{\rm fine}={e^2 v_x}/(4\pi\epsilon_0\epsilon_r\hbar v_y v_z)$ is the effective fine structure constant. For $q>q_c$ and for low energies, the novel gapless plasmon mode disperses as  
\bc
\omega_{\rm npl}^2\approx v_z^2 (\beta^2-1)^2G(\beta)~q_c (q-q_c)~.
\ec
The $\omega_{\rm npl}\propto (q-q_c)^{1/2}$ behaviour is consistent with the results from the hydrodynamic theory. 
The normal gapped plasmon mode in both type-I type-II DSM, has an anisotropic plasmon gap,  
owing to the anisotropic electronic dispersion. In a type-II DSM, the plasmon gap, $\omega_{\rm pl}(q=0)$, along the tilt axis is given by the root of the 
following transcendental equation
\bc
\omega^2 = \frac{\mu^2 v_z ^2}{\hbar^2  v_x^2}{\frac{4g_s\alpha_{\rm fine}}{3\pi\gamma(\omega)}}~, \hspace{.2cm}\gamma(\omega)=1+\frac{g_s\alpha_{\rm fine}v_z^2}{3\pi v_x^2}\ln\left|\frac{4\mathcal{E}_{\rm max}^2}{4\mu^2-\omega^2}\right|.
\ec
Here, the plasmon gap scales as $\omega_{\rm pl}(q=0) \propto \mu \propto n$, in contrast to the $\omega_{\rm pl} \propto n^{2/3}$ scaling in type-I DSM. 

The normal gapped plasmon mode was recently observed in PtTe$_2$ (a type-II DSM) along the direction perpendicular to the tilt axis ($\Gamma-K$) by means of high-resolution EELS (HREELS) \cite{PtTe2PRL,PhysRevB.99.045414}. HREELS analyzes the electrons reflected by the crystal surface with an energy resolution of a few meV \cite{PtTe2PRL,PhysRevB.99.045414}, and can transfer only momentum components parallel to the cleavage surface ($q_{\parallel}$) \cite{ibach1982electron}.  Therefore, plasmons along the tilt axis ($\Gamma-A$) are generally inaccessible to HREELS and also to other scattering techniques used to study the dispersion relation of low-energy collective modes, such as inelastic helium atom scattering (HAS) \cite{PhysRevLett.69.2951}. Conversely, momentum-resolved, energy-filtered transmission electron microscopy (EF-TEM) easily enables probing excitations along $\Gamma-A$. Unfortunately, the energy resolution of most EF-TEM apparatuses ($>200$ meV \cite{Egerton_2008}) is largely inadequate to detect gapless excitations. Nevertheless, recent technological advancements have been decisive to improve the energy resolution up to 18-50 meV \cite{Hageeaar7495,Ncom2041} with the next target to reach ~5 meV \cite{lovejoy_bacon_bleloch_dellby_hoffman_krivanek_2017}. Consequently, it is expected that in a few years the measurement of the dispersion relation of plasmonic modes along the tilt axis ($\Gamma-A$) will be experimentally feasible.

In summary, we predict a novel undamped gapless plasmon mode in a type-II DSM, arising from the presence of both electron and hole pockets at the Fermi energy. This novel mode exists beyond a critical wave-vector and only along the direction of the tilt axis which is usually the $C_n$ rotation axis in crystals hosting type-II DSM phase. Physically, it arises due to the 
out-of-phase oscillation of the density deviations in the electron and the hole pockets. 
A similar gapless (and possibly undamped) plasmon mode is also expected to arise in type-II WSM. 

\begin{acknowledgements}
AA and KS thank Sougata Mardanya and Barun Ghosh for stimulating discussions. 
\end{acknowledgements}

\appendix
  
\bibliography{refs}

\end{document}